\documentclass{article}

\pdfoutput=1
\usepackage{PRIMEarxiv}
\usepackage[utf8]{inputenc} 
\usepackage[T1]{fontenc}    
\usepackage[hidelinks,urlcolor=blue]{hyperref}
\usepackage{url}            
\usepackage{booktabs}       
\usepackage{amsfonts}       
\usepackage{nicefrac}       
\usepackage{microtype}      
\usepackage{lipsum}
\usepackage{fancyhdr}       
\usepackage{graphicx}       
\graphicspath{{media/}}     
\usepackage{textcomp}
\usepackage{natbib}
\usepackage{longtable}
\usepackage{xurl}
\usepackage[utf8]{inputenc}
\usepackage[T1]{fontenc}
\usepackage{graphicx}
\usepackage{amsmath}
\usepackage{float}

\begin{document}
\title{Why Companies ``Democratise'' Artificial Intelligence: The Case of Open Source Software Donations}

\author{
  Cailean Osborne \\
  Oxford Internet Institute \\
  University of Oxford \\
  Oxford, UK\\
  \texttt{cailean.osborne@oii.ox.ac.uk}
}

\maketitle

\begin{abstract}
Companies claim to ``democratise’’ artificial intelligence (AI) when they donate AI open source software (OSS) to non-profit foundations or release AI models, among others, but what does this term mean and why do they do it? As the impact of AI on society and the economy grows, understanding the commercial incentives behind AI democratisation efforts is crucial for ensuring these efforts serve broader interests beyond commercial agendas. Towards this end, this study employs a mixed-methods approach to investigate commercial incentives for 43 AI OSS donations to the Linux Foundation. It makes contributions to both research and practice. It contributes a taxonomy of both individual and organisational social, economic, and technological incentives for AI democratisation. In particular, it highlights the role of democratising the governance and control rights of an OSS project (i.e., from one company to open governance) as a structural enabler for downstream goals, such as attracting external contributors, reducing development costs, and influencing industry standards, among others. Furthermore, OSS donations are often championed by individual developers within companies, highlighting the importance of the bottom-up incentives for AI democratisation. The taxonomy provides a framework and toolkit for discerning incentives for other AI democratisation efforts, such as the release of AI models. The paper concludes with a discussion of future research directions.
\end{abstract}


\section{Introduction}
Companies are increasingly ``democratising'' artificial intelligence (AI). However, ``AI democratisation'' remains an ambiguous term, encompassing a variety of goals and methods \citep{seger_democratising_2023}, including the release of AI open source software (OSS) \citep{langenkamp_how_2022,srnicek_data_2022}, its donation to non-profit foundations \citep{yue_igniting_2024}, and the release of AI models (henceforth: open models) \citep{osborne_ai_2024,widder_open_2023}, among others. While press releases celebrate a myriad of benefits that AI democratisation promises for research and innovation, the commercial incentives driving such efforts are often obscured from public view. Given the ever-increasing impact of AI on society and the economy, understanding the commercial incentives for AI democratisation efforts is crucial so that we can ensure these efforts serve broader societal interests beyond commercial agendas.

Towards this end, this study presents an exploratory investigation of why companies democratise AI with a focus on OSS donations as one method, among others, of AI democratisation. Through a mixed-methods approach, combining the analysis of pre-donation technical pitches, post-donation blog posts, a questionnaire, and semi-structured interviews, it investigates commercial incentives for 43 AI OSS donations to the Linux Foundation (LF), making contributions to both research and practice. It makes contributions to both research and practice. It contributes a taxonomy of both individual and organisational social, economic, and technological incentives for AI democratisation. In particular, it highlights the role of democratising the governance and control rights of an OSS project (i.e., from one company to vendor-neutral, open governance) as a structural enabler for downstream goals, such as attracting external contributors, reducing development costs, and influencing industry standards, among others. Furthermore, it sheds light on the role of individual developers within companies, who champion and coordinate OSS donations, thus highlighting the relevance of the bottom-up incentives for AI democratisation. The taxonomy provides a framework and toolkit for discerning incentives for other AI democratisation efforts, such as the release of AI models \citep{white_model_2024}. 

The paper is structured as follows. First, it reviews prior work on the political economy of open source AI and OSS (Section~\ref{sec:donations-litreview}). Second, it presents the research aims and the study design (Section~\ref{sec:donations-studydesign}). Third, it reports the key findings (Section~\ref{sec:donations-results}). Fourth, it discusses the implications of the findings for research and practice as well as the threats to validity (Section~\ref{sec:donations-discussion}). Finally, the paper concludes with a summary of the key contributions (Section~\ref{sec:donations-conclusion}).
\section{Related Work}\label{sec:donations-litreview}

\subsection{``Democratising AI'': Narratives and Practices}

In the context of concerns about industry concentration and influence on AI research and development (R\&D) \citep{ahmed_growing_2023,kak_ai_2023,whittaker_steep_2021,varoquaux_hype_2024}, it has become \textit{en vogue} for companies to claim to ``democratise AI''---an altruism-laden term that is notoriously ambiguous. Prior work finds that it used as a catch-all term to encompass a variety of goals and practices \citep{seger_democratising_2023}, including the following:

\begin{itemize}
    \item \textbf{Democratising AI use:} Lowering entry barriers for the use of AI technologies, including but not limited to commercial products like OpenAI's ChatGPT or GitHub's Copilot, access to AI models through APIs or publicly available model weights, and the release of AI OSS like PyTorch and TensorFlow.
    \item \textbf{Democratising AI development:} Lowering the entry barriers for the development of AI technologies, including but not limited to the release of AI models and AI OSS.
    \item \textbf{Democratising AI profits:} Redistributing the economic value accrued to companies from their use of AI technologies to the respective users and impacted populations.
    \item \textbf{Democratising AI governance:} Distributing the decision-making power in the development or use of AI technologies to a wider community of stakeholders and impacted populations.
\end{itemize}

In most cases, AI democratisation is used to refer to the lowering of barriers for the use or the development of AI technologies, leading \citet{seger_democratising_2023} to conclude that, ``‘AI democratisation’ is a (mostly) unfortunate term''. Open source technologies and collaboration methods have been integral to AI democratisation efforts, offering the means of enabling both access to and participation in the development of AI. Commercial releases of AI OSS \citep{langenkamp_how_2022,srnicek_data_2022} and open models \citep{osborne_ai_2024,widder_open_2023} have contributed to the rapid growth of the open source AI ecosystem, which now comprises over 300 AI OSS libraries \citep{haddad_artificial_2022}, hundreds of thousands of open models \citep{osborne_ai_2024}, and over a million AI OSS repositories \citep{github_machine_2023}.

The prevalence of AI democratisation efforts begs the questions of why companies release their AI software and models, and what are the impacts thereof on the norms, practices, and potential trajectories of AI developer communities. Prior work hints at a number of incentives. Scholars contend that industry giants promote open source as an alternative to their concentrated power in the AI industry, whilst using it as a means to shape industry standards, benefit from user innovation, and ultimately extend their influence the norms and tools used by researchers and developers around the world \citep{widder_open_2023,srnicek_data_2022}. Nick Srnicek argues that ``the seemingly non-capitalist practice of releasing their AI software for free in fact obscures a significant capitalist battle between the major companies'' \citep{srnicek_data_2022}. This was evident in a leaked Google memo, which claimed that ``open source solutions will out-compete companies like Google or OpenAI'' and for this reason they should ``own the ecosystem and let open source work for us'' \citep{patel_google_2023}. As discussed below, this leaked memo highlights the ethical tensions that emerge from company’s attempts to exploit the collective efforts of OSS developer communities \citep{birkinbine_incorporating_2020,li_ethical_2022}.

Other companies, such as Meta, have been outspoken about the drivers of their open source AI strategy: by releasing AI software like PyTorch and large language models (LLM) like their LLaMA models, Meta seeks to increase adoption of its technology, improve their performance and safety through distributed feedback and innovation, and ultimately benefit from ecosystem effects. For example, upon releasing LLaMA 2,  Nick Clegg, Meta’s President of Global Affairs, explained that ``Openness isn’t altruism---Meta believes it’s in its interest. It leads to better products, faster innovation, and a flourishing market, which benefits us as it does many others'' \citep{clegg_nick_2023}. He argued that releasing LLaMA 2 would make it ``safer and better'' because it will benefit from the ``wisdom of the crowds.'' Clegg added that, ``A mistaken assumption is that releasing source code or model weights makes systems more vulnerable. On the contrary, external developers and researchers can identify problems that would take teams holed up inside company silos much longer.’’ Meanwhile Mark Zuckerberg, Meta's CEO, has explained publicly that Meta seeks to build an ecosystem around their AI software and models as a source of strategic advantage. For example, Zuckerberg explained to shareholders that the widespread use of PyTorch has ``been very valuable for us'' because it has facilitated their use  of external AI research and innovations that use PyTorch \citep{meta_meta_2023}. Furthermore, upon the release of LLaMA 3, he explained that they are not doing open source ``because we are, like, altruistic… I just want everyone to be using it because the more people who are using it, the more the flywheel will spin for making LLaMA better'' \citep{south_park_commons_mark_2024}. These statements provide some answers to why technology giants like Meta democratise AI, but it remains to be studied systematically why various kinds of companies, including startups, engage in AI democratisation efforts.

There are also concerns about ``open-washing'' by startups and industry giants alike, who have been promoting open models released under restrictive licenses and with limited transparency as ``open source'' \citep{white_model_2024,widder_open_2023,liesenfeld_opening_2023}. Liesenfeld and Dingemanse (\citeyear{liesenfeld_rethinking_2024}) argue that companies engage in open-washing to reap the benefits of open source (e.g., reputation rewards and adoption), whilst not actually complying with open source standards or norms (e.g., via restrictive licenses). For example, Meta released LLaMA 2 with much fanfare, claiming that the ``open source'' model would benefit research and innovation, its distribution under a novel license with restrictive commercial terms (i.e., any company with greater than 700 million monthly active users in the preceding month must request a license that Meta may grant in its sole discretion) received backlash from the open source community \citep{white_model_2024,widder_open_2023,maffulli_metas_2023}.  For example, Stefano Maffulli, the Executive Director of the Open Source Initiative, commented, ``Unfortunately, the tech giant has created the misunderstanding that LLaMA 2 is `open source' – it is not. 
Meta is confusing 'open source' with `resources available to some users under some conditions,' [which are] two very different things'' \citep{maffulli_metas_2023}.

Whether open source AI will deliver its promised equalising effects or lead to further industry concentration remains to be seen. However, it is important to view these open source AI developments in the context of developments and concentrations in the wider AI supply chain or ``AI stack'', including the hardware accelerators (i.e., chips), data, talent, and compute infrastructure required to develop and deploy AI systems in practice \citep{widder_open_2023,srnicek_data_2022,lehdonvirta_behind_2023,lehdonvirta_cloud_2023,don-yehiya_future_2024}. One may argue that  no matter how much AI software or how many AI models companies release publicly, regardless of their license choice, such AI democratisation efforts will do little to fundamentally reconfigure the distribution of power and resources in the wider AI industry.

This prior work provides insights into the open source AI strategies of industry giants and concerns related to open-washing. However, we still have significant gaps in our understanding of why and how different types of companies, beyond industry giants, employ open source as a means to ``democratise'' AI. To address this gap, in the next section, I draw on the extensive literature on the political economy of OSS, which provides a comprehensive theoretical foundation for understanding commercial incentives for AI democratisation efforts that are facilitated by open source technology.

\subsection{Commercial Incentives for OSS Development}

Companies have participated in the development of OSS in a myriad of ways since the late 1990s  \citep{broca_communs_2021,li_systematic_2024}, including by deploying developers to contribute to projects as part of their job responsibilities or corporate social responsibility initiatives \citep{dahlander_relationships_2005,dahlander_man_2006,lee_how_2015,zhang_companies_2021}, funding projects \citep{obrien_foss_2019,osborne_public-private_2024}, or joining  project steering committees \citep{butler_investigation_2018,wagstrom_vertical_2009}, among others. These are popular strategies through which companies seek to influence projects that develop maintain OSS that they use \cite{dahlander_man_2006,osborne_public-private_2024}. It is also common for companies to spin-out proprietary software as company-hosted OSS, where the host company controls the intellectual property of the project (e.g., by requiring contributors to sign a contributor license agreement) and employs the maintainers of the project \citep{zhou_inflow_2016,yue_igniting_2024}. This is a proven commercial strategy to increase adoption of their software, benefit from external contributions, win market share, or reduce a competitor’s market share \citep{west_contrasting_2005}. In some cases, a handful of companies share control of a project; for example, in 2017, Facebook and Microsoft jointly released the Open Neural Network Exchange (ONNX) to enable interoperability between various deep learning frameworks like PyTorch and TensorFlow \citep{candela_facebook_2017}.

An extensive literature discusses the diverse incentives for commercial  adoption and development of OSS at both the individual level (see Table~\ref{tab:OSS-donations-lit-incentives-ind}) and the organisational level (see Table~\ref{tab:OSS-donations-lit-incentives}). Bonaccorsi and Rossi’s (\citeyear{bonaccorsi_comparing_2006}) taxonomy of social, economic, and technological incentives at the individual and organisational levels provides an enduring framework for categorising these diverse incentives. In addition, they find that divergent incentives between individuals and organisations. While individuals are mostly driven by social and technological incentives, such as their personal interest \citep{benkler_wealth_2006,raymond_cathedral_2001,von_krogh_carrots_2012}, values \citep{kelty_two_2008,lakhani_how_2003,shah_motivation_2006}, or needs \citep{franke_satisfying_2003,hars_working_2002,roberts_understanding_2006}, companies are chiefly motivated by economic and technological factors, such as influencing industry standards \citep{lerner_incentives_2002,lindman_beyond_2009}, reducing development costs \citep{birkinbine_incorporating_2020,chesbrough_measuring_2023}, and recruiting external developers \citep{agerfalk_outsourcing_2008,fink_business_2003}. However, the incentives of individuals vary based on factors such as whether they are volunteers or paid \citep{lakhani_why_2003,dahlander_man_2006}, the governance structure of the OSS project \citep{shah_motivation_2006}, and their geography and cultural norms \citep{subramanyam_freelibre_2008,takhteyev_coding_2012}, among others. While individual developers, including volunteers, do not share the primary incentives of for-profit companies, they tend to accept commercial participation on condition that they comply with community norms \citep{bonaccorsi_comparing_2006}. Furthermore, commercial participation in OSS projects can also attract volunteers, who see their presence as a signal of the complexity of the project \citep{omahony_boundary_2008,santos_attraction_2013}. 

\begin{table*}[p]
    \centering
    \footnotesize
    \caption{Incentives for Individuals to Participate in OSS Development}
    \begin{tabular}{p{2cm}p{4cm}p{10cm}}        \toprule
        \textbf{Category} & \textbf{Sub-category} & \textbf{References} \\ \midrule
        \textbf{Social} 
        & Fun \& geek culture &  \citet{torvalds_just_2001, raymond_cathedral_2001,gerosa_shifting_2021,lakhani_why_2003,lakhani_how_2003,benkler_coases_2002,benkler_wealth_2006,hertel_motivation_2003,luthiger_pervasive_2007,roberts_understanding_2006,shah_motivation_2006,von_hippel_open_2003,xu_volunteers_2009,hemetsberger_fostering_2002}  \\ \\ 
        & Reputation \& peer recognition &  \citet{raymond_cathedral_2001,gerosa_shifting_2021,bezroukov_open_1999,ghosh_freelibre_2002,lakhani_why_2003,lakhani_how_2003,lerner_incentives_2002, fershtman_determinants_2004,ghosh_understanding_2007,ghosh_interviews_1998,hars_working_2002,hertel_motivation_2003,lattemann_framework_2005,lerner_simple_2002,okoli_investigating_2007,oreg_exploring_2008,roberts_understanding_2006,spaeth_communal_2008,stewart_impact_2006,xu_volunteers_2009,yu_knowledge_2007,hemetsberger_fostering_2002,loebbecke_open_2003,krishnamurthy_intrinsic_2006,howison_incentives_2013} 
 \\ \\
        & Altruism \& reciprocity &  \citet{raymond_cathedral_2001,gerosa_shifting_2021,raymond_homesteading_2001,bergquist_power_2001,bitzer_intrinsic_2007,crowston_open_2002,hars_working_2002,osterloh_trust_2003,david_community-based_2008,david_floss-us_2003,ghosh_open_2006,haruvy_harvesting_2003,lakhani_why_2003,oreg_exploring_2008,schofield_participation_2006,shah_motivation_2006,stewart_impact_2006,wu_empirical_2007,yu_knowledge_2007,nagle_report_2020,ding_towards_2023}
 \\ \\
        & Kinship with OSS community &  \citet{pfaff_open_1998,gerosa_shifting_2021,raymond_homesteading_2001,hars_working_2002,david_community-based_2008,hertel_motivation_2003,lakhani_how_2003,schofield_participation_2006,zeitlyn_gift_2003,bagozzi_open_2006,hemetsberger_fostering_2002}

 \\ \\
        & Ideology \& personal values &  \citet{stallman_gnu_1984,raymond_cathedral_2001,gerosa_shifting_2021,kelty_two_2008,takhteyev_coding_2012,schoonmaker_free_2018,david_community-based_2008,david_floss-us_2003,ghosh_open_2006,hertel_motivation_2003,lakhani_how_2003,shah_motivation_2006,stewart_impact_2006,xu_12_2006,yu_knowledge_2007,akiki_bigscience_2022} 
 \\ \midrule 
        \textbf{Economic}
        & Payment \& sponsorship & \citet{lerner_incentives_2002,feller_understanding_2002,hertel_motivation_2003,hars_working_2002,ghosh_open_2006,lakhani_why_2003,lattemann_framework_2005,luthiger_pervasive_2007,roberts_understanding_2006,howison_incentives_2013}
 \\ \\
        & Job-seeking \& career benefits &  \citet{lerner_open_2001,lerner_simple_2002,lakhani_release_2002,lakhani_why_2003,gerosa_shifting_2021,ghosh_freelibre_2002,subramanyam_freelibre_2008,hossain_regional_2021,hann_economic_2002,hars_working_2002,hertel_motivation_2003,riehle_economic_2007,roberts_understanding_2006,shah_motivation_2006,yu_knowledge_2007}
 \\ \\
        & Low opportunity costs for use
 &  \citet{bonaccorsi_altruistic_2003,kollock_economics_1999,lakhani_why_2003} \\ \midrule
        \textbf{Technological} 
        & Learning \& skill development
 &  \citet{bonaccorsi_comparing_2006, lakhani_why_2003,ghosh_freelibre_2002,david_community-based_2008,david_floss-us_2003,hars_working_2002,gerosa_shifting_2021,roberts_understanding_2006,oreg_exploring_2008,shah_motivation_2006,stewart_impact_2006,spaeth_communal_2008,von_hippel_open_2003,wu_empirical_2007,ye_toward_2003,hemetsberger_fostering_2002,ding_towards_2023}
 \\ \\
        & Use state-of-the-art software
 &  \citet{pfaff_open_1998,pavlicek_keys_2000,green_economics_2000,franke_satisfying_2003,feller_understanding_2002,ding_towards_2023}
 \\ \\
        & Personal need for software
 &  \citet{raymond_cathedral_2001,von_hippel_open_2003,von_hippel_innovation_2001,hars_working_2002,osterloh_trust_2003,bitzer_economics_2006,ghosh_freelibre_2002,hertel_motivation_2003,lattemann_framework_2005,schofield_participation_2006,shah_motivation_2006,kuan_open_2001,david_community-based_2008,ghosh_open_2006} 
 \\ \bottomrule
    \end{tabular}
    \label{tab:OSS-donations-lit-incentives-ind}

    \footnotesize \textit{N.B. This table was adapted from \citet{bonaccorsi_comparing_2006} and updated with more recent publications.}
    \end{table*}

\begin{table*}[p]
    \centering
    \small
    \caption{Incentives for Companies to Participate in OSS Development}
    \begin{tabular}{p{2cm}p{4cm}p{10cm}}        \toprule
        \textbf{Category} & \textbf{Sub-category} & \textbf{References} \\ \midrule
        \textbf{Social} & Conformity with OSS values & \citet{lerner_incentives_2002,osborne_public-private_2024} \\ \\ &
        Reciprocity to OSS community &  \citet{feller_understanding_2002,franck_reconciling_2002,osterloh_trust_2003,lerner_simple_2002,germonprez_open_2013} \\ \\ 
        & Reputation as OSS patron &  \citet{osterloh_trust_2003,lerner_simple_2002,pitt_penguins_2006} \\ \\ & 
        OSS values (strategy to reduce market power of competitors) & \citet{feller_understanding_2002} 
        
        \\ \midrule 
        \textbf{Economic} & Reduce development costs  &  \citet{agerfalk_outsourcing_2008,birkinbine_incorporating_2020,birkinbine_commons_2018,chesbrough_measuring_2023,feller_understanding_2002,hawkins_economics_2004,lindman_beyond_2009,markus_what_2000,marsan_adoption_2012,woods_open_2005,kendall_game_2016,germonprez_open_2013,lakhani_release_2002,linaker_how_2016,loebbecke_open_2003,nguyen_duc_coopetition_2017,nguyen-duc_software_2019,teixeira_understanding_2014,teixeira_collaboration_2014,teixeira_lessons_2015,li_ethical_2022} \\ \\ 
         & Independence from price and license policies of vendors &  \citet{chesbrough_measuring_2023,lerner_simple_2002} \\ \\
         & Revenue generation from complementary services \& products &  \citet{chesbrough_measuring_2023,feller_understanding_2002,lindman_beyond_2009,west_how_2003,wichmann_firms_2002}
         \\ \\
         & Recruit software engineers &  \citet{agerfalk_outsourcing_2008,fink_business_2003,lerner_simple_2002, mugrage_changing_2022,wichmann_firms_2002, marlow_activity_2013} \\ \\ 
         & Increase market competitiveness &  \citet{ahlawat_why_2021,chesbrough_measuring_2023,hossain_regional_2021,lindman_beyond_2009,loebbecke_open_2003, woods_open_2005, kendall_game_2016} \\ \midrule
        \textbf{Technological} & 
        Crowdsource innovation & \citet{agerfalk_outsourcing_2008,birkinbine_incorporating_2020,birkinbine_commons_2018,chesbrough_measuring_2023,feller_understanding_2002,hawkins_economics_2004,lindman_beyond_2009,markus_what_2000,marsan_adoption_2012,woods_open_2005,kendall_game_2016,germonprez_open_2013,lakhani_release_2002,linaker_how_2016,teixeira_understanding_2014,li_ethical_2022}
        \\ \\ &
        Enhance software security  &  \citet{alrawashdeh_user_2020,bonaccorsi_comparing_2006, birkinbine_incorporating_2020,chesbrough_measuring_2023,fink_business_2003,widder_open_2023, hecker_setting_1999,hawkins_economics_2004,henkel_selective_2006,von_hippel_open_2003,lerner_simple_2002,lindman_beyond_2009,krishnamurthy_cave_2005,germonprez_open_2013} \\ \\ 
         &         Increase adoption of software & \citet{lerner_simple_2002,wagstrom_vertical_2009,widder_open_2023,srnicek_data_2022} \\ \\ &
         Promote open standards \newline and technical interoperability &  \citet{chesbrough_measuring_2023,fink_business_2003,lerner_simple_2002, lindman_beyond_2009, wichmann_firms_2002, west_how_2003} \\ \\ 
         & Influence the technical roadmap \newline \& direction of OSS projects &  \citet{dahlander_man_2006,lakhani_why_2003,linaker_public_2020,lindman_beyond_2009,von_krogh_carrots_2012, kendall_game_2016}   \\ 
        \bottomrule
    \end{tabular}
    \label{tab:OSS-donations-lit-incentives}

    \footnotesize \textit{N.B. This table was adapted from \citet{bonaccorsi_comparing_2006} and updated with more recent publications.}
\end{table*}

For companies, economic and technological incentives are the most salient. In particular, the distributed production model of OSS development, involving many more skilled developers beyond those within the organisational boundaries of any single company, is viewed as means to decrease in-house R\&D costs \citep{lindman_beyond_2009,agerfalk_outsourcing_2008}. Benjamin Birkinbine argues that the greatest value of OSS for companies stems from the peer production model that expands the labour force that can test and develop the software. Specifically, he contends that the value for companies stems from the \textit{processes}, not the \textit{products}, of OSS development \citep{birkinbine_incorporating_2020}. These incentives were underscored in the aforementioned Google memo, which urged the company to ``own the ecosystem and let open source work for us'' \citep{patel_google_2023}. The extent of volunteer activity in OSS development, from bug-spotting to code contributions \citep{feller_understanding_2002,roberts_understanding_2006,zhou_inflow_2016}, raises ethical questions about the exploitation of volunteer work \citep{li_ethical_2022} and the failure of most companies to adequately reciprocate to support the sustainability of OSS projects \citep{birkinbine_incorporating_2020, osborne_public-private_2024}. For example, while Linus’ Law---i.e., that ``Given enough eyeballs, all bugs are shallow'' \citep{raymond_cathedral_2001}---is typically quoted to argue that the OSS development model offers security advantages over proprietary software development, one can extend it to convey the value of distributed bug-spotting and improvements that no single company must pay for by themselves. 
 
While volunteers make useful contributions to OSS projects \citep{barcomb_uncovering_2020, capiluppi_cathedral_2007,crowston_bug_2008,zhou_who_2015}, it is often the case that companies specifically seek to collaborate with other companies, including market rivals, as a means to jointly share R\&D costs \citep{zhang_how_2020,teixeira_collaboration_2014, teixeira_cooperation_2016,nguyen-duc_software_2019} and shape industry standards \citep{lerner_open_2001,lerner_simple_2002,west_contrasting_2005}. While inter-company collaborations do not necessarily exclude volunteers, it is not uncommon for companies to engage in strategic or contractual collaborations, involving private collaboration methods, which volunteers cannot participate in (Osborne et al, forthcoming). The prevalence of inter-company collaborations has turned many OSS communities ``from networks of individuals into networks of companies'' \citep{agerfalk_outsourcing_2008}, resulting in a tangle of cooperation and competition between companies that has become known as ``open source co-opetition’’ \citep{zhang_how_2020,teixeira_collaboration_2014, teixeira_cooperation_2016,nguyen-duc_software_2019}. Moreover, commercial participation in OSS development can help companies to improve their market position by undercutting the product of a market rival \citep{fink_business_2003,west_contrasting_2005} as well as to enhance their reputation as an OSS patron among developers \citep{feller_understanding_2002,bonaccorsi_comparing_2006,pitt_penguins_2006}, which in turn can help to recruit software developers \citep{agerfalk_outsourcing_2008,lindman_beyond_2009}.

The diversity of incentives of various stakeholders underlines the critical role of governance in OSS projects  \citep{omahony_boundary_2008}. Non-profit foundations have emerged as key mediators---or ``boundary organisations''---whose vendor-neutrality and open community governance have proven to be effective structural enablers of collaboration between ``unexpected allies'' \citep{omahony_boundary_2008}. For instance, the LF is reputed to facilitate ``communities of competitors,'' where ``market rivals...intentionally coordinate activities for mutual benefit in precise, market-focused, non-differentiating engagements'' \citep{germonprez_open_2013}. Foundations limit the dominance of any single company in OSS projects, which attracts new contributors to projects \citep{di_giacomo_key_2020,west_contrasting_2005,link_understanding_2016}, especially volunteers who are hesitant about performing free work for a company \citep{zhang_companies_2018,zhang_companies_2021,zhou_inflow_2016}, and increases their adoption \citep{zhou_inflow_2016}. However, foundations do not always prevent commercial dominance in OSS projects \citep{wagstrom_vertical_2009,zhang_companies_2018,zhang_corporate_2022}. For instance, around 10\% of companies account for 80\% of commits to projects in the OpenStack ecosystem \citep{zhang_companies_2021}. Moreover, governance changes resulting from donations do not guarantee activity increases. For example, a study of PyTorch's governance transition from Meta to the LF revealed no net increase in project activity and specifically that contributions from Meta decreased significantly, those from users (e.g., app developers and cloud providers) remained unchanged, but those from complementors (e.g., chip manufacturers) increased \citep{yue_igniting_2024}. While governance changes may address ``hold-up'' problems for certain companies, particularly for complementors whose value capture proposition depends on interoperability, they do not guarantee net increases in external contributions \citep{yue_igniting_2024}.

This review has provided a theoretical foundation for examining commercial incentives in AI democratisation efforts. The taxonomy of social, economic, and technological incentives at individual and organisational levels \citep{bonaccorsi_comparing_2006} offers a valuable framework for this study's exploratory aims. By applying this framework to AI OSS donations, this study aims to identify and categorise the commercial interests driving AI democratisation. The following section outlines the research aims and methodological approach in more detail.

\section{Study Design}\label{sec:donations-studydesign}
\subsection{Research Aims}
The objective of this study is to identify and categorise commercial interests for AI democratisation and thus contribute to advancing the nascent research agenda on the political economy of open source AI \citep{widder_open_2023,srnicek_data_2022, liesenfeld_rethinking_2024}. Specifically, it examines the following research question (RQ): Why do companies democratise AI? Given the various methods of AI democratisation \citep{seger_democratising_2023}, it focuses on AI OSS donations to foundations---that is, the transfer of an OSS project from a company’s ownership to a non-profit foundation \citep{omahony_emergence_2007}---which in the AI industry are commonly presented as acts of AI democratisation. While this narrow scope enables an in-depth analysis of one method of AI democratisation, it naturally limits the generalisability of the findings to others, such as the increasingly common releases of open models (see Section~\ref{donations-threats-to-validity}). 

Within this scope, a mixed-methods approach was employed to investigate the incentives for 43 OSS donations between May 2018 and October 2022 by a range of companies, from startups to multinational corporations, to the LF AI \& Data Foundation and PyTorch Foundation, two foundations under the LF that host OSS for data science and AI. The range of projects and companies form a diverse sample (see Table~\ref{tab:donations-projects}), accounting for various project maturity levels, company sizes, sectors, and countries \citep{runeson_guidelines_2008,easterbrook_selecting_2008}. Furthermore, the mixed-methods approach to multiple cases mitigates the uniqueness of single cases \citep{herriott_multisite_1983,eisenhardt_building_1989} and data sources or methods \citep{yin_case_2018,lehdonvirta_global_2019}, thus enhancing the validity of the findings. The study received ethical clearance by the University of Oxford CUREC review board prior to data collection.

\subsection{Case Presentation}
\subsubsection{LF AI \& Data Foundation}
The LF AI \& Data Foundation was founded in March 2018 as the LF Deep Learning Foundation and rebranded as the LF AI Foundation in May 2019, broadening its scope to encompass various AI sub-fields. In October 2020, it merged with the ODPi, an organisation promoting a big data software ecosystem. The foundation subsequently rebranded as the LF AI \& Data Foundation, acknowledging the vital role of data in AI R\&D. At the point of data collection (October 2022), the foundation had 51 member companies from North America, Europe, and East Asia. It hosted 42 OSS projects that had been donated by diverse organisations, including startups (e.g., AI Squared), research institutes (e.g., the Beijing Academy of AI), consortia (e.g. ONNX), management consultancies (e.g., McKinsey \& Co.), and technology giants (e.g., IBM, Samsung, and Tencent). When a company seeks to donate an OSS project to the LF AI \& Data Foundation, they must be a member organisation of the LF or be endorsed by a member and submit their proposal for review by the technical advisory council (TAC). The TAC comprises representatives from the various projects at the foundation and premier member companies, who vote on the approval of donation proposals. The LF AI \& Data Foundation segregates business and technical governance of hosted OSS projects, ensuring that developers retain technical control in their projects whilst the foundation assumes responsibility for funding, marketing, and license compliance, among others, and enforces open community governance \citep{dolan_how_2023}.

\subsubsection{PyTorch Foundation}
The PyTorch Foundation was established in September 2022 to host the popular PyTorch deep learning framework that had been donated by Meta \citep{pytorch_about_2023}. Its mission is to ``driv[e] the adoption of AI tooling by fostering and sustaining an ecosystem of open source, vendor-neutral projects integrated with PyTorch'' and ``to democratise state-of-the-art tools, libraries, and other components to make these innovations accessible to everyone'' \citep{pytorch_about_2023}. The PyTorch Foundation similarly maintains a separation between business and technical governance for the PyTorch project and wider ecosystem, with the PyTorch project retaining its technical governance structure while the foundation is responsible for funding, hosting expenses, and events, among others. The PyTorch Foundation manages the project’s assets, including its website, GitHub repository, and social media accounts, and enforces open community governance. Upon its launch, it formed a governing board comprising representatives from its initial members: AMD, Amazon Web Services, Google Cloud, Hugging Face, IBM, Intel, Meta, Microsoft, and Nvidia \citep{pytorch_about_2023}. The governing board members shape PyTorch's strategic direction through voting rights, contribute to the project's technical development and roadmap, and gain benefits such as early feature access and increased visibility in the PyTorch ecosystem, while being expected to actively support and promote PyTorch's growth and adoption.

\subsection{Data \& methods}

\subsubsection{Data Collection}
This study comprises two sources of primary data and two sourcs of secondary data. First, two types of secondary data were collected from the Internet for 43 AI OSS donations to the LF (42 donations to the LF AI \& Data Foundation and 1 donation to the PyTorch Foundation). First, pre-donation technical pitches to the TAC were collected from the LF AI \& Data Foundation wiki page \citep{lfaidata_lf_2023}. Second, post-donation blog posts by the LF and respective companies were collected from the LF AI \& Data Foundation website and through Google search queries in the format of “[company name] + [project name] + [LF AI \& Data Foundation]. This yielded 40 presentations (95\%) and 37 (88\%) blog posts for the 42 projects donated to the LF AI \& Data Foundation. It was not possible to collect a pre-donation technical pitch for the PyTorch donation because, as an inaugural project of its namesake foundation, it did not follow this process. Blog posts were accessed via the PyTorch Foundation website and Google search queries using the aforementioned format. A full list of the 43 OSS projects, donors, and respective document links is provided in Table~\ref{tab:donations-projects}.

\begin{figure*}[p]
  \centering
\includegraphics[width=0.89\textwidth,height=0.89\textheight]{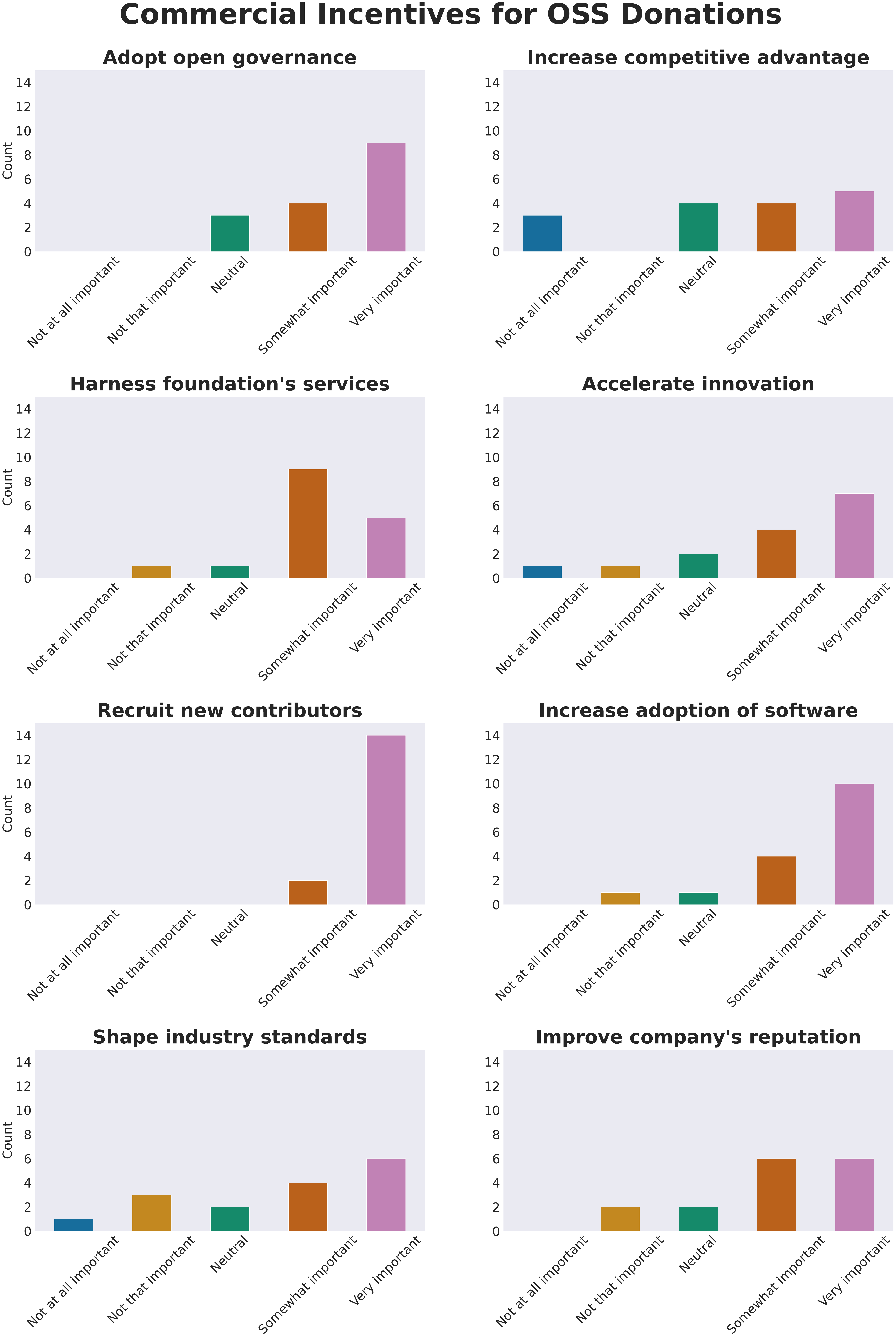}
\caption{Commercial Incentives for AI OSS Donations to the Linux Foundation}
  \label{fig:donations-incentives}

\footnotesize 
\textit{Sample: 16/43 (37\%) OSS projects hosted at LF AI \& Data Foundation and PyTorch Foundation in October 2022.}
\end{figure*}

Subsequently, primary data was collected through questionnaires and 12 semi-structured interviews with ten project maintainers who had donated the project and two foundation employees. First, a brief questionnaire was distributed to the maintainers to gather information on the donation process, incentives, and outcomes, as well as to recruit interviewees. It was distributed via the LF AI \& Data Foundation's mailing list and to the PyTorch maintainers via the Executive Director of the PyTorch Foundation, resulting in 16 responses from the maintainers of 16 projects at the LF AI \& Data Foundation and 0 responses from the maintainers of PyTorch (in total, 37\% of the 43 projects). Ten maintainers were recruited for interviews through the questionnaire, who worked for 9 companies, diverse by geography, size, and sector (see Table~\ref{tab:OSS-donations-interviewees}). In addition, the Executive Director of the foundations (N.B. same person) and a LF AI \& Data Foundation project coordinator were interviewed. The 12 interviews lasted between 30 and 60 minutes and were semi-structured, combining standardised questions about the donation process with tailored questions based on their questionnaire responses, adding depth to the quantitative findings in Figure~\ref{fig:donations-incentives}. The interviews were conducted digitally, and were recorded to aid analysis and to enhance the validity of the research findings \citep{yin_case_2018}.

\subsubsection{Data Analysis}\label{donations-thematic-analysis}
Thematic analysis was applied to systematically identify commonalities, patterns, and relationships in the qualitative data \citep{cruzes_recommended_2011}. A systematic six-step procedure was followed to enhance the reliability of this analysis \citep{braun_using_2006}. The initial coding procedure involved an integrated approach, combining the inductive coding \citep{charmaz_constructing_2006} and a deductive coding informed by prior work on incentives that exist at the levels of individual developers (see Table~\ref{tab:OSS-donations-lit-incentives-ind}) and organisations (see Table~\ref{tab:OSS-donations-lit-incentives}). This approach allowed for the identification of both commonalities with prior work and novel findings concerning open source AI democratisation efforts. The coding was conducted by the author until reaching saturation \citep{charmaz_constructing_2006}, then merged codes into 25 distinct themes (i.e., social, economic, and technological incentives at the level of individual developers and companies shown in Table~\ref{tab:donation-findings-taxonomy}). To address the limitation of single-author analysis, each step was rigorously documented, and the results were member-checked with the interviewees and discussed with two academic advisers  \citep{edwards_how_2013,king_interviews_2009}. Furthermore, the interviewees were invited to review the quotes attributed to their anonymised IDs and to state their attribution preference, ensuring consent for the inclusion of statements. Only three interviewees proposed revisions (e.g., to enhance specificity), indicating the resonance of the findings with the practitioners.

\begin{table}[h]
    \centering
    \small
    \caption{List of Interviewees \& Affiliations}
    \begin{tabular}{p{0.5cm}p{2.55cm}p{2.55cm}lp{3.5cm}}
        \textbf{ID} & \textbf{Project} & \textbf{Role}  &\textbf{Company Size} & \textbf{Sector} \\ \toprule
        A & BeyondML & Maintainer  &Small & Information technology \\ 
        B & Elyra & Maintainer  &Large & Information technology \\ 
        C & Elyra & Maintainer  &Large & Information technology \\ 
        D & Kedro & Maintainer  &Large & Professional services \\ 
        E & Kedro & Maintainer  &Large & Professional services \\ 
        F & KServe & Maintainer  &Large & Information technology \\ 
        G & Ludwig & Maintainer  &Small & Information technology \\ 
        H & NNstreamer & Maintainer  &Large & Information technology \\ 
        I & ONNX-MLIR & Maintainer  &Large & Information technology \\ 
        J & ONNX-MLIR & Maintainer  &Medium & Information technology \\ 
        K & LF AI \& Data & Project coordinator  &Medium & Non-profit foundation \\
        L & LF AI \& Data, \newline PyTorch Foundation & Executive director  &Medium & Non-profit foundation \\ \bottomrule
    \end{tabular}
    \label{tab:OSS-donations-interviewees}
\end{table}

\subsubsection{Reflexivity}\label{sec:donations-reflexivity}
In social science research, it is critical that one as a researcher engages in critical self-evaluation of the one’s positionality and disiplinary conventions, and how they may influence one’s research, from its initial design through to the reporting of and presentation of  research findings \citep{finlay_outing_2002}. The exercise of reflexivity was particularly important for the credibility of this study, given the author’s affiliation with the LF as a research contractor. A social identity map was employed as a tool to encourage reflection on positionality and to address potential biases in three areas \citep{jacobson_social_2019}. First, at the outset, it was used to consider the effects of the LF affiliation on data access and potential threats to reproducibility, recognising that the LF affiliation likely increased the willingness of foundation staff and project maintainers to participate in the study. To enhance reproducibility, the data collection strategy relied primarily on publicly available information (e.g., mailing lists) and information sheets sent with invitations explicitly stated the independent purpose and funding for the research. Second, to minimise social desirability bias in interview responses, the research's independent purpose and funding mentioned in the information sheet were explained to interviewees at the beginning of each interview. Third, a journal was maintained during the thematic analysis to document coding choices, and the findings were shared with two academic advisers to review the interpretations. While these actions contribute to enhancing the credibility and reliability of the research process, it must be acknowledged that it is difficult, if not impossible, to perform bias-free research and, therefore, it should be understood as an imperfect, yet best-effort attempt by the author to control for and mitigate potential biases in the research process \citep{jacobson_social_2019}.

\section{Results}\label{sec:donations-results}
The findings reveal an interplay of social, economic, and technological incentives at both individual and organisational levels for OSS donations to foundations (see Table~\ref{tab:donation-findings-taxonomy}). This section presents these findings, contributing a more nuanced understanding of the bottom-up and top-down incentives behind AI democratisation efforts by companies.

\subsection{Individual-level Incentives}

Individual employees often play a crucial role in championing and coordinating the donation process within their companies. 38\% of questionnaire respondents stated that the decision to donate was driven from the bottom up by developers, while 13\% of respondents stated  the decision was made by individuals who held both developer and managerial roles (e.g., startup founders). These findings underscore the importance of understanding individual incentives in shaping commercial decisions to donate OSS projects to foundations.

\subsubsection{Social Incentives}

At the individual level, social incentives are as a significant driver for OSS donations. Two key themes stood out: reciprocity and personal reputation. Many developers expressed a deep-seated ethos of `giving back'' to the community that supports their work. For instance, Respondent A (BeyondML) stated:

\begin{quote}
\textit{The vast majority of proprietary models and software in data science and machine learning are built on open source, so being part of that and contributing to that is really important to me personally and to our company.}
\end{quote}

This sentiment was echoed by Respondent H (NNStreamer), who noted that their donation was ``just for our own satisfaction'' as OSS users and developers. These responses highlight the personal investment many developers have in the OSS ecosystem and their desire to contribute to its growth and sustainability. Personal reputation is also a significant social incentive, with successful donations to major foundations representing an opportunity for individual developers to enhance their standing in the OSS community. For example, Respondent J (ONNX-MLIR) explained that having a project accepted by a major foundation provides credibility among peers and brings developers like himself closer to their ``dream of having your big open source project with 1000s of stars.'' Respondent I (ONNX-MLIR) added that improving one's reputation also leads to career benefits, as one can be hired based on one's reputation, and that individuals' aspirations align with company's goals to improve their corporate reputation as an OSS-friendly workplace.  

\subsubsection{Economic Incentives}

Two primary economic incentives at the individual level are career benefits and access to foundation support services.The reputational gains from OSS contributions often translate into tangible career benefits. For example, respondent I (ONNX-MLIR) noted that achievements in OSS projects make developers more competitive in the AI job market, as their expertise at the intersection of software engineering and AI becomes both known and knowable in the wider open source AI ecosystem. This enhanced visibility and credibility in the job market creates a powerful incentive for individuals to support OSS donations. As mentioned above, Respondent I (ONNX-MLIR) pointed out that these  incentives of individual developers also align with organisational goals of fostering a skilled workforce, attracting and recruiting competitive AI talent, and improving their corporate reputation among OSS developers.

The desire to harness foundation support services is another important economic incentive. Many developers viewed foundation support as a means to address challenges faced by maintainers in managing projects alongside their full-time employment. 87\% of respondents claimed this support was important for them (see Figure~\ref{fig:donations-incentives}). Respondent F (KServe) elaborated on this point:

\begin{quote}
\textit{We, as developers, don't have a lot of time for [outreach and marketing]. We sought to benefit from support services, such as outreach to new contributors and marketing.}
\end{quote}

Similarly, Respondent H (NNStreamer) sought marketing support to increase project visibility and to attract external developers. These examples illustrate how foundations are perceived to provide resources that individual maintainers often lack or cannot secure within their own companies, which in turn helps to enhance the growth of their OSS projects.

\subsubsection{Technological Incentives}

Technological incentives at the individual level include ensuring project sustainability and enabling the use of collaboration tools. Project sustainability is a significant concern for maintainers. Respondent G (Ludwig) provided a compelling example, describing how he donated his project to ensure its survival following organisational restructuring and his personal departure from the company. He viewed the transition to a foundation as an effective strategy to ensure the project's continuation. This case demonstrates how personal attachment to projects, which he described as his ``baby'', can drive individuals to seek sustainable governance models for their OSS projects, especially when their affiliation with the company comes to an end. It also highlights the use of OSS donations as a mechanism to preserve source code that might otherwise be lost due to corporate changes or neglect. 

The ability to use preferred collaboration tools is another technological incentive. Respondents D and E (Kedro) explained that transitioning the project to the LF AI \& Data Foundation made it easier to use tools like Slack and Discord, which were either forbidden or difficult to get approval for at their company. As Respondent E (Kedro) explained, ``It untied our hands from our own bureaucracy.'' They elaborated that this freedom from corporate constraints concerning collaboration tools not only enhances developer productivity and satisfaction but also aligns with broader organisational goals of fostering innovation and efficiency, thus representing a win-win scenario for them.

\subsection{Organisational-level Incentives}

While individual employees play a crucial role in championing OSS donations within their companies, organisational incentives (i.e., corporate strategies) ultimately matter above all in the decision-making process, with 44\% of respondents stating that the donation was a top-down decision by managers and 13\% respondents stating that the decision was made by individuals who held both developer and managerial roles within their company.

\subsubsection{Social Incentives}

At the organisational level, three primary social incentives emerged: adopting open governance, reciprocating to the OSS ecosystem, and building the company's reputation. The adoption of open governance upon donating a project to a  foundation is a salient incentive, with 81\% of respondents reporting it as important. This change in governance model is viewed as a structural enabler for downstream goals. For example, Detakin's press release highlighted this incentive:

\begin{quote}
\textit{The LF AI \& Data provides a vendor-neutral governance structure that can help the project grow broad industry collaboration. Even more importantly, becoming a LF AI \& Data project ensures that OpenLineage can never belong to a company.}
\end{quote}

Similarly, Lyft emphasised the importance of a ``neutral holding ground'' when donating both Amundsen and Flyte. These statements underscore the perceived value of neutral governance in fostering collaboration and ensuring the independence of projects.

Similar to individuals, reciprocity to the OSS ecosystem is a key social incentive at the organisational level. Several respondents cited the critical importance of OSS dependencies in their companies' proprietary products and services, and perceived the donation of their OSS project as one way to ``give back'', as explained by Respondent A (BeyondML). In a similar vein, Respondent C (Elyra) underscored the impact of OSS for advances in AI as a reason for why he champions giving back to the ecosystem:

\begin{quote}
\textit{The democratisation of AI software is really what is helping industry advance. If you look back like 10-20 years ago, it was very hard and you needed to have a specific set of skills to be able to even build a very simple model. Today with all the tools and stuff, it's much easier.}
\end{quote}

Respondent C (Elyra) explained that their team's decision to donate Elyra stemmed in part from their desire to play their part in advancing industry. This sentiment was echoed in several companies' post-donation press releases, framing their donations as acts of AI democratisation. For instance, Uber stated that by donating Pyro, it hoped ``to facilitate greater opportunities for researchers worldwide and [to] make deep learning and Bayesian modelling more accessible.''

Building corporate reputation emerged as the third significant social incentive at the organisational level. Many respondents explained they hoped that the reputation of the LF AI \& Data would enhance their company's credibility in the open source AI ecosystem, with 75\% of respondents reporting it as important. This incentive was particularly strong for startups and companies without an established reputation in AI. Respondent A (BeyondML) explained:

\begin{quote}
\textit{One of the things that we hoped to get from the LF is its name recognition, obviously just about every developer in the world knows about the LF or knows the term Linux. So having that kind of badge, if you will, immediately gives you a level of credibility with your project.}
\end{quote}

Respondents D and E (Kedro) echoed this sentiment, stating in their pre-donation pitch that they ``would like to leverage the initial marketing announcements to build credibility in their technical and product-related capability''. Subsequently, in the post-donation press release, their company stated:

\begin{quote}
\textit{It's a substantial step forward for our organisation on our open source journey. The consultancy sector has traditionally been highly protective of intellectual property, but it's clear that open, collaborative innovation will help power the next phase for analytics technology.}
\end{quote}

These findings highlight three key social incentives driving organisational decisions to donate AI OSS projects. The adoption of open governance emerges as a key structural enabler for fostering collaboration and ensuring project independence. Reciprocity to the OSS ecosystem reflects companies' recognition of their dependence on open source technologies and their desire to contribute to industry advancement. Additionally, the opportunity to enhance corporate reputation, particularly for less established companies, serves as a salient motivator. These social incentives are strategically important for companies seeking to position themselves favourably within the open source AI ecosystem.

\subsubsection{Economic Incentives}

Several economic incentives at the organisational level inform the decision to donate OSS projects, including attracting external contributors, reducing development costs, diversifying project funding, and harnessing foundation support services. Indeed, being able to attract new contributors to their project is a critical incentive for OSS donations, with 100\% respondents reporting it as important. Respondent J (ONNX-MLIR) described OSS donations as a strategic trade-off, where companies exchange full control of their OSS project for the aspired for benefits of distributed development involving a community of contributors. He noted the self-interested logic:

\begin{quote}
\textit{We continue doing the same work as we would if it wasn't an open source project, but there's this expectation that we're going to benefit from a community helping us achieve our own goals.}
\end{quote}

Respondent G (Ludwig) explained that above all it is beneficial for attracting contributors from other companies, who ``prefer not to contribute to projects that are started by companies that could be competitors. They don't trust it as fully open source if it was started by Uber, Google, Facebook, or whatever company.'' Meanwhile, Respondent B (Elyra) provided a more nuanced view:

\begin{quote}
\textit{In reality, in all the projects that I've seen, they are still driven by the main inventors. Open governance just means that the feedback comes from additional contributors ... It's more like a community project.}
\end{quote}

However, respondents cautioned that changing the governance model does not guarantee more or useful external contributions. Respondent D (Kedro) explained she had rejected pull requests ``because they were not up to scratch'', while Respondent G (Ludwig) discussed the role of mentorship in training external developers to become effective contributors, which he highlighted requires the company to invest time and money.

Diversification of funding is another significant economic incentive. For example, respondent E (Kedro) discussed the relevance of client concerns about the financial dependence on the host company:

\begin{quote}
\textit{Clients would ask, 'What will happen if your team does not exist tomorrow?' They were afraid that if we left the code, they wouldn't be able to get new versions and then it would become unmaintained.}
\end{quote}

Respondent D (Kedro) speculated that the only reason they were able to convince their senior management to approve the open-sourcing of Kedro was due to this client pressure. In addition, some companies seek to attract or increase external investment in their project. Several respondents saw this as a primary reason for Meta's donation of PyTorch, speculating that members of the newly founded governing board of the PyTorch Foundation would likely invest more resources in PyTorch than they did before.

Harnessing foundation support services also emerged as a significant economic incentive at the organisational level, with 87\% of respondents reporting it as important. For example, Respondent A (BeyondML) described the resources and infrastructure provided by foundations as ``stability offerings'' that could help scale and sustain projects beyond the means of their startup, thus making OSS donations an attractive option for sustainable project growth.

\subsubsection{Technological Incentives}

At the organisational level, several technological incentives were identified, including ecosystem integration and adoption, software improvements, faster innovation, and influencing industry standards. Respondents highlighted that joining a foundation offered ecosystem benefits. For example, respondent C (Elyra) noted:

\begin{quote}
\textit{Together with [open governance], we thought being in an ecosystem of other machine learning and AI projects would foster collaboration and integration, exposing Elyra to more use cases.}
\end{quote}

This perspective illustrates how companies view foundation ecosystems as platforms for enhanced collaboration and project visibility, potentially leading to more adoption and more diverse applications of their software. This was confirmed by 88\% of respondents who reported increasing adoption as an important incentive for their company.

Software improvements and faster innovation were also identified as significant technological incentives. Respondent B (Elyra) observed that ``That's something which is one strength of open source; it's much more properly tested than [proprietary] software.'' An additional incentive is the aim to accelerate development, implicitly thanks to new contributors who would join the project, with 69\% of respondents reporting it as important. This underlines the aforementioned perspective of Respondent J (ONNX-MLIR) that AI democratisation is in large part self-interested, using open source as a means to reduce internal costs, speed up innovation, and enhance software quality.

Influencing industry standards emerged as another key technological incentive, with 63\% of respondents reporting it as important for their company. A number of respondents speculated that the ambition behind Meta's donation of PyTorch was to make it the open standard for deep learning, bringing an end to the age-old rivalry between PyTorch and Google's TensorFlow. Respondent E (Kedro) commented:

\begin{quote}
\textit{I think that move was actually made to destroy TensorFlow because TensorFlow [does not have] open governance.}
\end{quote}

Respondent D (Kedro) suggested that Meta was seeking to beat TensorFlow out of the market since Google would struggle to compete with the strategic alliance of industry giants that had formed under the PyTorch Foundation's governing board (which even includes Google). Respondent B (Elyra) was more direct, describing it as ``a death knell to TensorFlow.'' Reflecting on their experience of overseeing donations to the foundation, Respondent L (LF) explained:

\begin{quote}
\textit{Every company may have a different set of incentives but what's common across all of them is the desire to make sure that the project becomes successful in the long term and becomes the de facto project for its given functionality.}
\end{quote}

These findings reveal a complex interplay of social, economic, and technological incentives driving companies to donate AI OSS to foundations. The incentives span both individual and organisational levels, highlighting the multifaceted nature of AI democratisation efforts. These insights provide a nuanced understanding of the strategies and considerations that shape commercial decisions, practices, and strategies in the open source AI ecosystem.

\section{Discussion}\label{sec:donations-discussion}

\subsection{Implications for Research}\label{sec:donations-impl-research}

\subsubsection{Disambiguating ``AI Democratisation'' Narratives}
The findings shed light on the inconsistent use of ``AI democratisation'' in commercial narratives surrounding OSS donations, extending the work of \citet{seger_democratising_2023}. They show that in the context of OSS donations, the most relevant interpretation is the democratisation of governance, as companies transfer already open-sourced software from their single-vendor governance to open governance provided by vendor-neutral foundations. This transition of governance and control rights, however, is not an end in itself but a strategy that aims to realise goals, such as increasing adoption and recruiting new contributors to the project. It is often the companies seek, in particular, participation by other companies, who previously would hesitate to contribute to a company's OSS project, confirming prior findings on ``hold-up'' problems \citep{yue_igniting_2024}. These findings challenge the notion that AI democratisation efforts are primarily altruistic. OSS donations are a strategic calculation, where companies relinquish control of their OSS project in exchange for assumed benefits. This perspective aligns with and extends the work of Widder et al. \citet{widder_open_2023} and Srnicek \citet{srnicek_data_2022}, who argue that open source practices in AI serve commercial interest. This should not be surprising, as for-profit companies are not charities, but it provides an empirical evidence base to illustrate that ``there is no such thing as a free lunch'' and to understand the possible objectives behind such AI democratisation efforts. Moreover, the analysis reveals the role of individual developers who advocate for OSS donations within their companies, demonstrating that donations result from an interplay between both individual and organisational incentives. 

\subsubsection{A Taxonomy for Understanding AI Democratisation Incentives}
The findings can be summarised in a taxonomy of commercial incentives for AI OSS donations (see Table~\ref{tab:donation-findings-taxonomy}). Drawing on the taxonomy by Bonaccorsi and Rossi (\citeyear{bonaccorsi_comparing_2006}, see Tables~\ref{tab:OSS-donations-lit-incentives-ind}-\ref{tab:OSS-donations-lit-incentives}), this taxonomy categorises incentives at both individual and organisational levels, grouped into social, economic, and technological categories. It provides a framework for understanding the intersecting bottom-up and top-down incentives behind OSS donations, demonstrating that they are driven by a combination of individual and organisational factors. The value of this taxonomy lies in its ability to provide a structured approach to analysing the incentives for AI democratisation efforts, offering researchers and practitioners a tool to systematically examine the incentives at play in different scenarios.

The findings corroborate prior work on the salience of economic and technological incentives for companies \citep{bonaccorsi_comparing_2006}, with the democratisation of governance in large part used as a means to various economic and technological ends, such as recruiting external developers \citep{agerfalk_outsourcing_2008,fink_business_2003}, reducing development costs \citep{birkinbine_incorporating_2020,chesbrough_measuring_2023}, and influencing industry standards \citep{lerner_incentives_2002,lindman_beyond_2009}. According to the questionnaire, the most important incentive for companies is to recruit new contributors to the project, who can help improve the quality and competitiveness of the software, ultimately serving the interests of the donor company. As Respondent J (ONNX-MLIR) noted, ``there's an expectation that we're going to benefit from a community helping us achieve our own goals.'' These findings corroborate statements by Big Tech companies about their open source AI strategies, who want to ``own the ecosystem and let open source work for us'' \citep{patel_google_2023} and for ``everyone to be using [LLaMA] because the more people who are using it, the more the flywheel will spin'' \citep{south_park_commons_mark_2024}.  This aligns with ethical concerns about commercial exploitation of volunteer work in OSS without adequately reciprocating to the OSS ecosystem \citep{li_ethical_2022,birkinbine_incorporating_2020}. 

However, the findings also add nuance to this perspective. While the promise of being able to attract new contributors to their project is indeed important for many companies, it is not universally paramount. For some companies, particularly those with substantial resources or an existing community, other incentives such as standard-setting take precedence \citep{lerner_incentives_2002,lindman_beyond_2009,widder_open_2023}. The case of PyTorch, which was donated by Meta and already had a large contributor community, exemplifies this scenario. Even for companies primarily seeking to attract new contributors, the findings show that these benefits are far from guaranteed. The interviews identified two key challenges. First, the mere act of donation does not ensure that a project will attract new contributors. Second, even if a project can recruit new contributors, the quality of their contributions may not meet the project's standards. Companies reject external contributions that ``are not up to scratch'' (Respondent D, Kedro) and need to invest in mentoring and training resources for external contributors (Respondent G, Ludwig). Companies may need to invest considerable time and resources into community building and contributor development, challenging the notion of open source as a straightforward cost-saving measure. This finding provides important nuance to the ``free labour'' thesis, and provides avenues for future research into the real costs and benefits of spinning-out or donating OSS projects.

It is also worth noting that the relative importance of these incentives vary depending on the context, company size, and type of the AI technology  being ``democratised'' (e.g., OSS, AI model, or other). While this taxonomy provides a starting point for understanding commercial incentives for AI democratisation, it should be used as a flexible framework rather than a definitive list. Future research should build on it by testing its applicability to different approaches to AI democratisation, such as AI model and dataset releases \citep{seger_democratising_2023,seger_open-sourcing_2023,don-yehiya_future_2024}. On the one hand, incentives such as building a company's reputation, increasing adoption, recruiting new contributors, reducing development costs, and shaping industry standards likely apply to AI model releases. On the other hand, the adoption of open governance is, at least to this day, specific to the context of OSS donations to foundations. To date, while companies have released models, they have not transferred them to a vendor-neutral foundation for open governance. This may in part be due to the lack of established practices for openly governing AI models given their relative complexity compared to OSS in terms of the resources required to develop and maintain AI models and their constituent components \citep{white_model_2024}, or in part because the incentives may be more similar to those for spinning-out software, such as increasing adoption \citep{west_contrasting_2005}. At this stage, we can only speculate and, therefore, the commercial incentives and strategies for AI model releases, including governance choices, as well as their implications for norms and practices in the open source AI ecosystem, warrant further investigation. 

\begin{table}[t]
    \centering
    \small
    \caption{Commercial Incentives for AI OSS Donations}
    \begin{tabular}{p{2.5cm}p{4.8cm}p{4.8cm}} 
        \toprule
        \textbf{Category} & \textbf{Individual Incentives} & \textbf{Organisational Incentives} \\ \midrule
        \textbf{Social} & 
        
            • Personal interests in AI and OSS \newline
            • Reciprocate to OSS ecosystem \newline
            • Build personal reputation 
         & 
            • Transition to open governance \newline
            • Reciprocate to community/ecosystem \newline
            • Build company's reputation \newline
            • Join community of companies
         \\ \midrule
        
        \textbf{Economic} & 
        • Professional and career benefits \newline
        • Harness foundation services
         &
            • Recruit new contributors to project \newline
            • Reduce development costs \newline
            • Diversify project funding \newline
            • Harness foundation services \newline
            • Recruit talent to join company \newline
            • Increase competitive advantage
         \\ \midrule
        
        \textbf{Technological} & 
            • Ensure project sustainability \newline
            • Increase visibility and adoption \newline
            • Contribute to open source AI stack \newline
            • Access to new collaboration tools
         & 
            • Increase visibility and adoption \newline 
            • Accelerate innovation \newline
            • Improve software quality \newline
            • Foster technical interoperability \newline
            • Shape industry standards
         \\ \bottomrule
    \end{tabular}
    \label{tab:donation-findings-taxonomy}
\end{table}

\subsubsection{From Intentions to Outcomes: Evaluating the Long-term Impact of OSS Donations}
The findings capture the stated incentives of companies at a particular point in time. However, the actual realisation of these incentives and goals may vary and would require longitudinal study to verify. For instance, the study by \citet{yue_igniting_2024} on PyTorch's governance transition provides a compelling case study that both supports and challenges some of the incentives identified in this research. Their finding that the governance change led to increased participation from complementors (e.g., chip manufacturers) aligns with the incentive of recruiting new contributors identified in this study. However, the decrease in contributions from Meta following the transition suggests that the realisation of incentives may be more complex than initially anticipated. This complexity is further underscored by the fact that users (e.g., app developers and cloud providers) did not change their rate of participation, indicating that different stakeholders may respond differently to governance changes. These insights call for more longitudinal studies that track the outcomes of OSS donations over time. Such research could help verify whether long-term economic and technological benefits are indeed realised. Furthermore, it could shed light on how the balance of contributions shifts between the donor company and external contributors following a governance change. The PyTorch case also highlights the importance of considering the specific type of external contributors when evaluating the success of OSS donations. The differential response between complementors and users suggests that future research should delve deeper into how different types of stakeholders perceive and respond to changes in project governance. Future research could explore how companies balance the potential benefits of increased external contributions against the possible reduction in their own incentives to invest in the project. While this study provides a comprehensive taxonomy of incentives for OSS donations, the emerging evidence on the impacts of such donations highlights the need for further research.

\subsection{Implications for Practice}
The findings have several important implications for practitioners in open source AI development and governance. First, the findings underscore the need for greater transparency in corporate communications about AI democratisation efforts. The inconsistent use of ``AI democratisation'' underlines that practitioners should be more specific and transparent about their goals, as suggested by \citet{seger_democratising_2023}. Companies could clearly communicate the specific aspects of AI they aim to democratise (e.g., access, development, or governance) and explain how their actions contribute to these goals. Furthermore, greater transparency about the benefits that companies seek to gain from their democratisation efforts could foster more trust in corporate-community relations. This transparency could lead to more realistic expectations from the open source community and potentially more sustainable collaborations. 

Second, the taxonomy provides a toolkit for researchers and practitioners to understand the various social, economic, and technological incentives that may be driving AI democratisation efforts. It is also useful in shedding light on relevant incentives at both the individual developer level and the organisational level, revealing that both bottom-up and top-down incentives may be at play when companies announce they are democratising AI. Furthermore, practitioners can use this toolkit to examine how well commercial incentives align with the broader interests of the open source AI community, which may help companies and volunteers identify areas of synergy and enable more ethical collaboration. 

Third, the findings highlight the importance of foundations that facilitate collaboration between diverse stakeholders. While the benefits of foundations for developing and governing OSS are by now well understood, to this day companies are not transferring AI models to foundations for open governance. As discussed above, this may in part be due to the lack of established practices for openly governing AI models given their relative complexity in terms of the resources required to develop and maintain AI models and their various constituent components \citep{white_model_2024}. A question for practitioners to explore is whether, and if so how, the foundation-hosting and open governance model could be adopted for the hosting and governance of AI models. To answer this question, AI companies should engage foundations, not just as potential hosts of models, but as active partners in shaping the future of open source AI development.

By employing the toolkit and implementing these practices, practitioners can work towards a more transparent and sustainable approach to AI democratisation that acknowledges commercial interests while fostering the broader goals of open and collaborative AI development. This approach could lead to more ethical and effective AI technologies that better serve societal needs while still allowing for innovation and commercial success.

\subsection{Threats to Validity}\label{donations-threats-to-validity}
The validity of the findings are evaluated with reference to guidance for case study research and qualitative research methods in software engineering research \citep{runeson_guidelines_2008,yin_case_2018, easterbrook_selecting_2008}.

\subsubsection{Credibility}
Credibility refers to the believability of the findings  \citep{easterbrook_selecting_2008}. The primary threat stems from the author's affiliation with the LF as a research contractor, creating potential conflicts of interest and research biases, despite the independent purpose and funding of this research study. As explained in Section~\ref{sec:donations-reflexivity}, a social identity map was used to identify and address potential biases in three areas: data access, collection, and analysis \citep{jacobson_social_2019}. As previously stated, while a number of step were taken to enhance the credibility of the research process, it is acknowledged that there is nonetheless a risk of biases that were not controlled for or mitigated and as such it should be understood as an imperfect but best-effort attempt by the author. Another challenge concerned gaining access to data about commercial strategies. The working solution involved drawing on four sources---pre-donation technical pitches, post-donation press releases, questionnaires, and interviews---which offered a triangulated account of incentives \citep{yin_case_2018}. While this approach provided four perspectives, it was inevitably constrained by the limited participation of company-affiliated developers, who were willing or allowed to participate in a research interview, and by the fact that some strategic incentives may never be shared publicly. In addition, the willingness of maintainers to complete the questionnaire and/or participate in an interview may have been affected by company policies (e.g., NDAs). 

\subsubsection{Robustness}
Robustness concerns strength, reliability, and soundness of the study's design, methods, and findings. There was a risk of social desirability bias or response bias in the interviews, owing to the author's affiliation with the LF. Steps were taken to minimise the risk of these biases, such as proactively communicating the independent purpose and funding of this study in the interview invitations and at the beginning of every interview. Another threat stems from the thematic analysis. The author sought to maximise the robustness of this analysis by following best-practice guidelines \citep{braun_using_2006} and integrating approaches to coding qualitative data \citep{cruzes_recommended_2011}. In addition, the structure of the taxonomy poses risks to the validity of the findings. In some cases, it was difficult to demarcate incentives at the two units of analysis, which may have led to a misclassification or oversimplification. That being said, efforts were made to develop the categories by thoroughly reviewing prior work and member-checking findings with interviewees to evaluate their accuracy and resonance with practitioners \citep{lincoln_naturalistic_1985}.

\subsubsection{Transferability}
Transferability concerns the generalisability of qualitative research findings. There are two key threats to generalisability. First, the narrow focus on OSS donations as a method of AI democratisation limits the generalisability of the taxonomy, which both includes aspects specific to OSS donations (e.g., democratising governance or foundation support) and excludes aspects that may be typical of other methods of AI democratisation (e.g., open model releases). Second, this study may be limited by the specific characteristics of the sample of companies, OSS projects, and the foundations. The sample was drawn from the LF AI \& Data Foundation and PyTorch Foundation, two foundations under the LF, which may not fully represent other foundations. Future research directions were recommended in Section~\ref{sec:donations-impl-research} to advance our understanding of the commercial incentives for different approaches to AI democratisation.

\subsubsection{Dependability}
Dependability refers to the consistency of the research process. A comprehensive list of secondary documents collected for each OSS project is provided in Table~\ref{tab:donations-projects}. These documents were triangulated by questionnaires and interviews with maintainers and staff from the foundations. With their consent, all interviews were recorded and transcribed to aid the analysis \citep{yin_case_2018}. With regards to the data analysis, common guidelines were followed for the systematic analysis of qualitative data \citep{braun_using_2006,cruzes_recommended_2011}. Additionally, the accuracy of the analysis was validated through member-checking findings with interviewees, ensuring their resonance with practitioners.

\section{Conclusion}\label{sec:donations-conclusion}
Companies are ``democratising'' AI through various approach and with various goals. While these efforts tend to be celebrated for facilitating science and innovation, the strategic incentives driving these apparently altruistic acts are in large part hidden from public view. However, as the development and application of AI technologies has an ever-increasing impact on society and the economy, it becomes imperative to better understand the incentives and strategies for AI democratisation as well as whose interests AI democratisation ultimately serves. Towards this end, this study employed a mixed-methods approach to investigate the commercial incentives for 43 AI OSS donations to the LF. The findings highlight an interplay of  social, economic, and technological incentives at both individual and organisational levels. In the case of OSS donations, the democratisation of governance is treated as a social means for primarily economic and technological ends, such as attracting external contributors, reducing development costs, and influencing industry standards, among others. Furthermore, the study illustrates the role of individual developers, who champion OSS donations within their companies, thus highlighting the relevance of the bottom-up incentives for AI democratisation. While some incentives are unique to OSS donations, the taxonomy serves both as a theoretical foundation and as a practical toolkit for understanding the commercial incentives for related AI democratisation efforts, such as AI model releases. As the number of open models continues to grow at a rapid pace, it is timely for researchers to turn their attention to the commercial incentives and strategies driving these AI democratisation efforts, as well as the effects thereof on the norms, practices, and potential trajectories of the open source AI ecosystem at large.

\newpage
\section*{Acknowledgements}
The author thanks all research subjects for their participation in this study.

\section*{Declarations}
\subsection*{Funding}
This study was funded by the UK Economic and Social Research Council (Grant Number: ES/P000649/1).

\subsection*{Competing interests}
The author was affiliated with the LF as a research contractor during the research period. However, the study was independently funded by the UK Economic and Social Research Council (Grant Number: ES/P000649/1). 

\subsection*{Ethics approval}
This study obtained ethical clearance from University of Oxford’s Research Ethics Committee prior to data collection.

\subsection*{Consent}
As per the ethics approval requirements at the University of Oxford, all research subjects were asked to provide informed consent prior to data collection and were informed of their rights to withdraw their data at any stage. In addition, research subjects confirmed the affiliations and quotes included in this paper prior to its submission for publication. 

\subsection*{Data and/or Code availability}
N/A

\subsection*{Authors' contributions}
All contributions were made by the author and any errors made are his. 

\newpage
\bibliographystyle{apalike}
\bibliography{manuscript.bib}

\newpage
\appendix
\section{OSS Donations to LF AI \& Data Foundation and PyTorch Foundation}\label{donations-appendix}

\begin{longtable}{p{2.5cm}p{2.5cm}p{2cm}p{4cm}p{4cm}}
    \caption{OSS Donations to the LF AI \& Data and PyTorch Foundations} 
    
    \label{tab:donations-projects} \\
    \toprule
    \footnotesize \textbf{Project} & \footnotesize \textbf{Donor} & \footnotesize \textbf{Date} & \footnotesize \textbf{TAC Proposal} & \footnotesize \textbf{Press Release} \\ \midrule
    \endfirsthead 
    \toprule
    \textbf{Project} & \footnotesize \textbf{Company} & \footnotesize \textbf{Date} & \footnotesize \textbf{TAC Proposal} & \footnotesize \textbf{Press Release} \\ \midrule
    \endhead   \footnotesize  Acumos & \footnotesize AT\&T, \newline Tech Mahdra & \footnotesize 2018-05 & \footnotesize 
    \footnotesize \url{https://drive.google.com/file/d/1L6fFhZnFqeR3mwy8Ya/KVoRGzCUjdnyrM} & \footnotesize  \url{https://www.acumos.org/news/2018/11/14/lf-deep-learning-delivers-first-acumos-ai-release-making-it-easier-to-deploy-and-share-artificial-intelligence-models/} \\ \addlinespace \footnotesize 
        Angel & \footnotesize Tencent & \footnotesize 2018-08 & \footnotesize  \url{https://wiki.lfaidata.foundation/pages/viewpage.action?pageId=7733341\&preview=/7733341/18481216/GMT20191121-140452\_LF-AI-Foun\_1920x1080.mp4} & \footnotesize  \url{https://www.linuxfoundation.org/press/press-release/lf-deep-learning-adds-two-new-framework-projects-to-expand-community-and-ecosystem} \\ \addlinespace \footnotesize 
        Egeria & \footnotesize IBM, ING & \footnotesize 2018-08 & \footnotesize  \url{https://wiki.lfaidata.foundation/pages/viewpage.action?pageId=7733341\&preview=/7733341/30408948/September\%2024\%2C\%202020\_LF\%20AI\%20TAC\%20Deck\%202.pptx} & \footnotesize  \url{https://www.linuxfoundation.org/press/press-release/new-ai-data-foundation-combines-industrys-fastest-growing-open-source-developments-in-artificial-intelligence-and-open-data} \\ \addlinespace \footnotesize 
        Elastic \newline Deep \newline Learning & \footnotesize Baidu & \footnotesize 2018-08 & \footnotesize N/A & \footnotesize  \url{https://www.linuxfoundation.org/press/press-release/lf-deep-learning-adds-two-new-framework-projects-to-expand-community-and-ecosystem} \\ \addlinespace \footnotesize 
        Horovod & \footnotesize Uber & \footnotesize 2018-12 & \footnotesize  \url{https://wiki.lfaidata.foundation/pages/viewpage.action?pageId=7733341\&preview=/7733341/30408797/August\%2013\%2C\%202020\_LF\%20AI\%20TAC\%20Deck\%20-\%20compressed2.pptx} & \footnotesize  \url{https://www.uber.com/en-GB/blog/horovod-deep-learning-foundation/} \\ \addlinespace \footnotesize 
        Pyro & \footnotesize Uber & \footnotesize 2019-01 & \footnotesize  \url{https://drive.google.com/file/d/1B0ZkJUKVZoJxsaUkge/02kGKddiZX8DLj/view} & \footnotesize  \url{https://www.uber.com/en-GB/blog/pyro-lf-deep-learning-foundation/} \\ \addlinespace
        \footnotesize Adlik & \footnotesize ZTE & \footnotesize 2019-09 & \footnotesize  \url{https://wiki.lfaidata.foundation/pages/viewpage.action?pageId=7733341\&preview=\%2F7733341\%2F12091694\%2FTAC+recording+08152019.mp4} & \footnotesize  \url{https://lfaidata.foundation/blog/2019/10/21/lf-ai-welcomes-adlik-as-newest-incubation-project/} \\ \addlinespace \footnotesize 
        ONNX & \footnotesize ONNX \newline Community & \footnotesize 2019-11 & \footnotesize  \url{https://wiki.lfaidata.foundation/pages/viewpage.action?pageId=7733341\&preview=/7733341/18481160/TAC\%2010-24-2019.mp4} & \footnotesize  \url{https://cloudblogs.microsoft.com/opensource/2019/11/14/onnx-joins-linux-foundation/} \\ \addlinespace \footnotesize 
        Marquez & \footnotesize WeWork & \footnotesize 2019-12 & \footnotesize  \url{https://wiki.lfaidata.foundation/pages/viewpage.action?pageId=7733341\&preview=/7733341/18481417/TAC-12192019.mp4} & \footnotesize N/A \\ \addlinespace \footnotesize 
        sparklyr & \footnotesize RStudio & \footnotesize 2019-12 & \footnotesize  \url{https://wiki.lfaidata.foundation/pages/viewpage.action?pageId=7733341\&preview=/7733341/18481269/TAC-recording-12052019.mp4} & \footnotesize  \url{https://lfaidata.foundation/blog/2020/01/29/sparklyr-joins-lf-ai-as-its-newest-incubation-project-scaling-data-science-and-machine-learning-workflows-using-apache-spark-and-r/} \\ \addlinespace \footnotesize 
        Milvus & \footnotesize Zilliz & \footnotesize 2020-01 & \footnotesize  \url{https://wiki.lfaidata.foundation/pages/viewpage.action?pageId=7733341\&preview=/7733341/22249639/January\%2016\%2C\%202020\_LF\%20AI\%20TAC\%20Deck.pdf} & \footnotesize  \url{https://www.prnewswire.com/news-releases/milvus-the-ai-search-engine-originally-developed-by-zilliz-joins-lf-ai-as-new-incubation-project-301038716.html} \\ \addlinespace \footnotesize 
        OpenDS4All & \footnotesize IBM, ODPi, UPenn & \footnotesize 2020-02 & \footnotesize  \url{https://wiki.lfaidata.foundation/pages/viewpage.action?pageId=7733341\&preview=/7733341/30408948/September\%2024\%2C\%202020\_LF\%20AI\%20TAC\%20Deck\%202.pptx} & \footnotesize  \url{https://community.ibm.com/community/user/ai-datascience/blogs/ana-echeverri1/2020/02/28/opends4all-is-live} \\ \addlinespace \footnotesize 
        NNStreamer & \footnotesize Samsung & \footnotesize 2020-03 & \footnotesize  \url{https://wiki.lfaidata.foundation/pages/viewpage.action?pageId=7733341\&preview=/7733341/24281106/March\%2012\%2C\%202020\_LF\%20AI\%20TAC\%20Deck\_v2.pdf} & \footnotesize  \url{https://research.samsung.com/news/LF-AI-Foundation-Announces-NNStreamer-as-Its-Newest-Incubation-Project} \\ \addlinespace \footnotesize  
        ForestFlow & \footnotesize DreamWorks \newline Animation & \footnotesize 2020-04 & \footnotesize  \url{https://wiki.lfaidata.foundation/pages/viewpage.action?pageId=7733341\&preview=/7733341/24281142/March\%2026\%2C\%202020\_LF\%20AI\%20TAC\%20Deck.pdf} & \footnotesize  \url{https://research.dreamworks.com/dreamworks-animation-releases-forestflow-machine-learning-model-server-to-the-open-source-community/} \\ \addlinespace \footnotesize 
        Ludwig & \footnotesize Uber & \footnotesize 2020-05 & \footnotesize  \url{https://wiki.lfaidata.foundation/pages/viewpage.action?pageId=7733341\&preview=/7733341/24281544/Ludwig\%20LFAI.pdf} & \footnotesize  \url{https://www.uber.com/blog/introducing-ludwig/} \\ \addlinespace \footnotesize 
        Adversarial Robustness Toolbox & \footnotesize IBM & \footnotesize 2020-06 & \footnotesize  \url{https://wiki.lfaidata.foundation/pages/viewpage.action?pageId=7733341\&preview=/7733341/30409001/October\%208\%2C\%202020\_LF\%20AI\%20TAC\%20Deck.pdf} & \footnotesize  \url{https://developer.ibm.com/open/centers/codait/trusted-ai/} \\ \addlinespace \footnotesize 
        AI Explainability 360 & \footnotesize IBM & \footnotesize 2020-06 & \footnotesize  \url{https://wiki.lfaidata.foundation/pages/viewpage.action?pageId=7733341\&preview=/7733341/30409001/October\%208\%2C\%202020\_LF\%20AI\%20TAC\%20Deck.pdf} & \footnotesize  \url{https://developer.ibm.com/open/centers/codait/trusted-ai/} \\ \addlinespace \footnotesize 
        AI Fairness 360 & \footnotesize IBM & \footnotesize 2020-06 & \footnotesize  \url{https://wiki.lfaidata.foundation/pages/viewpage.action?pageId=7733341\&preview=/7733341/30409001/October\%208\%2C\%202020\_LF\%20AI\%20TAC\%20Deck.pdf} & \footnotesize  \url{https://developer.ibm.com/open/centers/codait/trusted-ai/} \\ \addlinespace \footnotesize 
        Amundsen & \footnotesize Lyft & \footnotesize 2020-07 & \footnotesize  \url{https://wiki.lfaidata.foundation/pages/viewpage.action?pageId=7733341} & \footnotesize  \url{https://eng.lyft.com/amundsen-1-year-later-7b60bf28602} \\ \addlinespace \footnotesize 
        DELTA & \footnotesize DiDi & \footnotesize 2020-09 & \footnotesize  \url{https://wiki.lfaidata.foundation/pages/viewpage.action?pageId=7733341\&preview=/7733341/30408797/August\%2013\%2C\%202020\_LF\%20AI\%20TAC\%20Deck\%20-\%20compressed2.pptx} & \footnotesize  \url{https://lfaidata.foundation/blog/2021/06/17/delta-joins-lf-ai-data-as-new-incubation-project/} \\ \addlinespace \footnotesize 
        Feast & \footnotesize Gojek & \footnotesize 2020-09 & \footnotesize  \url{https://wiki.lfaidata.foundation/pages/viewpage.action?pageId=7733341\&preview=/7733341/30408948/September\%2024\%2C\%202020\_LF\%20AI\%20TAC\%20Deck\%202.pptx} & \footnotesize  \url{https://feast.dev/blog/feast-joins-the-linux-foundation-for-ai-data/} \\ \addlinespace \footnotesize 
        SOAJS & \footnotesize Herron Tech & \footnotesize 2020-09 & \footnotesize  \url{https://wiki.lfaidata.foundation/download/attachments/7733341/September\%2010\%2C\%202020\_LF\%20AI\%20TAC\%20Deck\%20-\%20updated.pdf?version=1\&modificationDate=1599741898000\&api=v2} & \footnotesize N/A \\ \addlinespace \footnotesize 
        DataPractices & \footnotesize DataPractices Org & \footnotesize 2020-12 & \footnotesize  \url{https://wiki.lfaidata.foundation/download/attachments/7733341/October\%205\%2C\%202020\_LF\%20AI\%20TAC\%20Deck\%282\%29.pdf?version=1\&modificationDate=1604497702000\&api=v2} & \footnotesize N/A \\ \addlinespace \footnotesize 
        JanusGraph & \footnotesize JanusGraph \newline Community & \footnotesize 2021-01 & \footnotesize  \url{https://wiki.lfaidata.foundation/download/attachments/7733341/December\%203\%2C\%202020\_LF\%20AI\%20TAC\%20Deck.pdf?version=2\&modificationDate=1606864208000\&api=v2} & \footnotesize  \url{https://lfaidata.foundation/blog/2021/01/12/janusgraph-joins-lf-ai-data-as-new-incubation-project/} \\ \addlinespace \footnotesize 
        Flyte & \footnotesize Lyft & \footnotesize 2021-02 & \footnotesize  \url{https://wiki.lfaidata.foundation/download/attachments/7733341/February\%2025\%2C\%202021\_LF\%20AI\%20TAC\%20Deck.pdf?version=1\&modificationDate=1614021188000\&api=v2} & \footnotesize  \url{https://eng.lyft.com/flyte-joins-lf-ai-data-48c9b4b60eec} \\ \addlinespace \footnotesize 
        Datashim & \footnotesize IBM & \footnotesize 2021-03 & \footnotesize  \url{https://wiki.lfaidata.foundation/download/attachments/7733341/January\%2014\%2C\%202021\_LF\%20AI\%20TAC\%20Deck.pdf?version=1\&modificationDate=1610631576000\&api=v2} & \footnotesize  \url{https://lfaidata.foundation/blog/2021/03/23/datashim-joins-lf-ai-data-as-new-incubation-project/} \\ \addlinespace \footnotesize 
        RosaeNLG & \footnotesize BNP Paribas & \footnotesize 2021-03 & \footnotesize  \url{https://wiki.lfaidata.foundation/pages/viewpage.action?pageId=7733341\&preview=/7733341/39092353/March\%2011\%2C\%202021\_LF\%20AI\%20TAC\%20Deck(2).pdf} & \footnotesize  \url{https://lfaidata.foundation/blog/2021/04/28/rosaenlg-joins-lf-ai-data-as-new-sandbox-project/} \\ \addlinespace \footnotesize 
        Substra & \footnotesize OWKIN & \footnotesize 2021-03 & \footnotesize  \url{https://wiki.lfaidata.foundation/download/attachments/7733341/March\%2025\%2C\%202021\_LF\%20AI\%20TAC\%20Deck.pdf?version=1\&modificationDate=1616590543000\&api=v2} & \footnotesize  \url{https://lfaidata.foundation/blog/2022/11/28/owkin-launches-open-science-push-by-open-sourcing-ai-software-substra-and-releasing-two-open-source-ai-innovations-at-neurips/} \\ \addlinespace \footnotesize 
        Kompute & \footnotesize The Institute \newline for Ethical AI & \footnotesize 2021-05 & \footnotesize  \url{https://wiki.lfaidata.foundation/download/attachments/7733341/May\%206\%2C\%202021\_LF\%20AI\%20TAC\%20Deck.pdf?version=2\&modificationDate=1620316765000\&api=v2} & \footnotesize  \url{https://lfaidata.foundation/blog/2021/08/26/kompute-joins-lf-ai-data-as-new-sandbox-project/} \\ \addlinespace \footnotesize 
        OpenLineage & \footnotesize Datakin, IBM & \footnotesize 2021-07 & \footnotesize  \url{https://wiki.lfaidata.foundation/download/attachments/7733341/December\%2015\%202022\_LF\%20AI\%20TAC\%20Deck.pdf?version=1\&modificationDate=1673523471000\&api=v2} & \footnotesize  \url{https://openlineage.io/blog/joining-lfai/} \\ \addlinespace \footnotesize 
        TonY & \footnotesize LinkedIn & \footnotesize 2018-09 & \footnotesize  \url{https://wiki.lfaidata.foundation/pages/viewpage.action?pageId=7733341\&preview=/7733341/43287136/July\%2015\%2C\%202021\_LF\%20AI\%20TAC\%20Deck.pdf} & \footnotesize  \url{https://engineering.linkedin.com/blog/2018/09/open-sourcing-tony--native-support-of-tensorflow-on-hadoop} \\ \addlinespace \footnotesize 
        Kedro & \footnotesize McKinsey \newline QuantumBlack & \footnotesize 2021-08 & \footnotesize  \url{https://wiki.lfaidata.foundation/download/attachments/7733341/August\%2026\%2C\%202021\_LF\%20AI\%20TAC\%20Deck.pdf?version=1\&modificationDate=1630089002000\&api=v2} & \footnotesize  \url{https://medium.com/quantumblack/kedro-joins-the-linux-foundation-to-become-an-open-standard-for-machine-learning-engineering-b0061152ff73} \\ \addlinespace \footnotesize 
        KServe & \footnotesize KServe \newline Community & \footnotesize 2021-11 & \footnotesize  \url{https://wiki.lfaidata.foundation/download/attachments/7733341/October\%2018\%2C\%202021\_LF\%20AI\%20TAC\%20Deck.pdf?version=1\&modificationDate=1637936838000\&api=v2} & \footnotesize  \url{https://lfaidata.foundation/blog/2022/02/24/kserve-joins-lf-ai-data-as-new-incubation-project/} \\ \addlinespace \footnotesize 
        OpenBytes & \footnotesize Graviti & \footnotesize 2021-11 & \footnotesize  \url{https://wiki.lfaidata.foundation/download/attachments/7733341/October\%2021\%2C\%202021\_LF\%20AI\%20TAC\%20Deck.pdf?version=1\&modificationDate=1635325659000\&api=v2} & \footnotesize  \url{https://www.linuxfoundation.org/press/press-release/linux-foundation-and-graviti-announce-project-openbytes-to-make-open-data-more-accessible-to-all} \\ \addlinespace \footnotesize 
        Artigraph & \footnotesize Replica & \footnotesize 2022-01 & \footnotesize  \url{https://wiki.lfaidata.foundation/download/attachments/7733341/January\%2027\%2C\%202022\_LF\%20AI\%20TAC\%20Deck.pdf?version=1\&modificationDate=1643715756000\&api=v2} & \footnotesize  \url{https://lfaidata.foundation/uncategorized/2022/04/13/lf-ai-data-announces-artigraph-as-new-sandbox-project/} \\ \addlinespace \footnotesize 
        1chipML & \footnotesize Ericsson & \footnotesize 2022-04 & \footnotesize  \url{https://wiki.lfaidata.foundation/download/attachments/7733341/April\%207\%2C\%202022\_LF\%20AI\%20TAC\%20Deck\%20\%281\%29.pdf?version=1\&modificationDate=1650500517000\&api=v2} & \footnotesize  \url{https://lfaidata.foundation/blog/2022/07/21/lf-ai-data-announces-three-new-sandbox-projects/} \\ \addlinespace \footnotesize 
        BeyondML & \footnotesize Squared AI & \footnotesize 2022-06 & \footnotesize  \url{https://wiki.lfaidata.foundation/pages/viewpage.action?pageId=7733341\&preview=/7733341/61964452/June\%2016\%2C\%202022\_LF\%20AI\%20TAC\%20Deck.pdf} & \footnotesize  \url{https://lfaidata.foundation/blog/2022/07/21/lf-ai-data-announces-three-new-sandbox-projects/} \\ \addlinespace \footnotesize 
        FlagAI & \footnotesize BAAI & \footnotesize 2022-06 & \footnotesize  \url{https://wiki.lfaidata.foundation/download/attachments/7733341/June\%2030\%2C\%202022\_LF\%20AI\%20TAC\%20Deck.pdf?version=1\&modificationDate=1657325720000\&api=v2} & \footnotesize N/A \\ \addlinespace \footnotesize 
        FedLCM & \footnotesize VMWare & \footnotesize 2022-08 & \footnotesize  \url{https://wiki.lfaidata.foundation/download/attachments/7733341/July\%2028\%2C\%202022\_LF\%20AI\%20TAC\%20Deck\%20\%281\%29.pdf?version=1\&modificationDate=1660155848000\&api=v2} & \footnotesize  \url{https://blogs.vmware.com/opensource/2022/10/27/open-source-project-fedlcm-to-the-lf-ai-data/} \\ \addlinespace \footnotesize 
        FATE & \footnotesize LinkedIn & \footnotesize 2022-08 & \footnotesize  \url{https://wiki.lfaidata.foundation/download/attachments/7733341/August\%2025\%2C\%202022\_LF\%20AI\%20TAC\%20Deck.pdf?version=1\&modificationDate=1662390660000\&api=v2} & \footnotesize  \url{https://cloudblogs.microsoft.com/opensource/2022/09/12/feathr-feature-store-joins-lf-ai-data-foundation/} \\ \addlinespace \footnotesize 
        OpenDataology & \footnotesize OpenDataology Community & \footnotesize 2022-08 & \footnotesize  \url{https://wiki.lfaidata.foundation/download/attachments/7733341/August\%2011\%2C\%202022\_LF\%20AI\%20TAC\%20Deck.pdf?version=1\&modificationDate=1661373254000\&api=v2} & \footnotesize N/A \\ \addlinespace \footnotesize 
        PyTorch & \footnotesize Meta & \footnotesize 2022-09 & \footnotesize N/A & \footnotesize  \url{https://pytorch.org/blog/PyTorchfoundation/} \\ \addlinespace \footnotesize 
        Elyra & \footnotesize IBM & \footnotesize 2022-10 & \footnotesize  \url{https://wiki.lfaidata.foundation/pages/viewpage.action?pageId=7733341\&preview=/7733341/39092353/March\%2011\%2C\%202021\_LF\%20AI\%20TAC\%20Deck(2).pdf} & \footnotesize  \url{https://developer.ibm.com/blogs/open-source-elyra-ai-toolkit-simplifies-data-model-development/} \\ \bottomrule
    \label{tab:OSS-donations-all-projects}
\end{longtable}

\end{document}